# Estimate of the Globing Warming Caused by the Retreat of Polar Sea Ice

*Alfred Laubereau and Hristo Iglev*

Physik-Department E11, Technische Universität München, James-Franck-Strasse, D-85748 Garching, Germany

ABSTRACT.

The growing concentrations of the greenhouse gases $CO_2$, $CH_4$ and $N_2O$ (GHG) in the atmosphere are often considered as the dominant cause for the global warming during the past decades. The reported temperature data however do not display a simple correlation with the concentration changes since 1880 so that other reasons are to be considered to contribute notably. An important feature in this context is the shrinking of the polar ice caps observed in recent years. We have studied the direct effect of the loss of global sea ice since 1955 on the mean global temperature estimating the corresponding decrease of the terrestrial albedo. Using a simple 1-dimensional model the global warming of the surface is computed that is generated by the increase of GHG and the albedo change. A modest effect by the GHG of 0.08 K is calculated for the period 1880 to 1955 with a further increase by 0.18K for 1955 to 2015. A larger contribution of 0.55 ± 0.05 K is estimated for the melting of polar sea ice (MSI) in the latter period, i.e. it notably exceeds that of the GHG and may be compared with the observed global temperature rise of 1.0 ± 0.1 K during the past 60 years. Our data also suggest a delayed response of the mean global temperature to the loss of sea ice with a time constant of approximately 20 years. The validity of the theoretical model and the interrelation between GHG-warming and MSI-effect are discussed.

**KEYWORDS** albedo, greenhouse effect

INTRODUCTION:

The surface temperature of the earth depends on various natural and anthropogenic factors that effect the solar input and the infrared reemission of the earth. Examples are volcanic activities, oceanic temperature profiles and the concentration of the greenhouse gases (1–5). $CO_2$ is an important factor because of its pronounced concentration rise during the past 140 years, mostly due to human activities. Previous work considering elaborate models for the climate zones of the earth arrived at the conclusion that the gas is responsible for more than 50% of the warming effect, while methane and $N_2O$ were also proposed to be quite relevant (6–10). On the other hand, analyses of black carbon and tropospheric ozone reveal that anthropogenic pollutants also contribute notably (11). For black carbon a contribution second after $CO_2$ was reported (12). The importance of snow cover and ice extent in the Northern Hemisphere was recognized by various authors leading to a positive feedback of surface reflectivity on climate and was quantified from seasonal observations (13,14). Other effects were also discussed, e.g. the remarkable correlation between global temperature and sunspot activities that is not well understood at the present time (15).

In this paper we wish to quantify the role of the polar ice for the mean surface temperature, $T_{surf}$, of the earth starting from reported ice data. Satellite observations on the maximum and minimum sea ice areas are available since 1979. Fig. 1a) shows those data (black points/Arctic and red points/Antarctic) including linear fits (16–18). In contrast to the MSI of the Arctic the recent increase of sea ice in the Antarctic should be noted. Calculated annual averages are indicated by dashed lines for the northern (black) and southern sea ice area (red), respectively. The data are extended back to 1955 by the help of auxiliary information on the sea ice extent (19–21). The latter observations are summarized in Fig. 1b). Since the northern sea ice extent (black circles and black solid line) displays an approximately linear decrease since 1955 the same behavior is taken for the northern sea area in Fig. 1a) (dashed black line); i.e. the linear slope for the measurements 1979 – 2015 is extrapolated to earlier years. For Antarctic sea ice the situation is different. A decrease of the southern sea ice extent is indicated by the data in Fig. 1b) for 1955 to 1979 (red triangles and solid red lines in the Fig.) followed by a small increase, as also found for the ice area in Fig. 1a). Assuming that the ratio between Antarctic sea ice extent and sea ice area is approximately constant from 1955 to 1979, the annual mean area of Antarctic sea ice is estimated from the extent data (dotted red line in Fig. 1a).

The data of Fig. 1 suggest a loss of sea ice area of respectively 28% and 23% in the Arctic and Antarctic since 1955. The values are used below to estimate a reduction of the albedo factor of approximately 0.8 ± 0.1 % because of the much smaller backscattering of sea water compared to snow-covered ice for incident sun light. A smaller reflection from the surface corresponds to an enlarged net input of solar radiation, i.e. increase of $T_{surf}$.

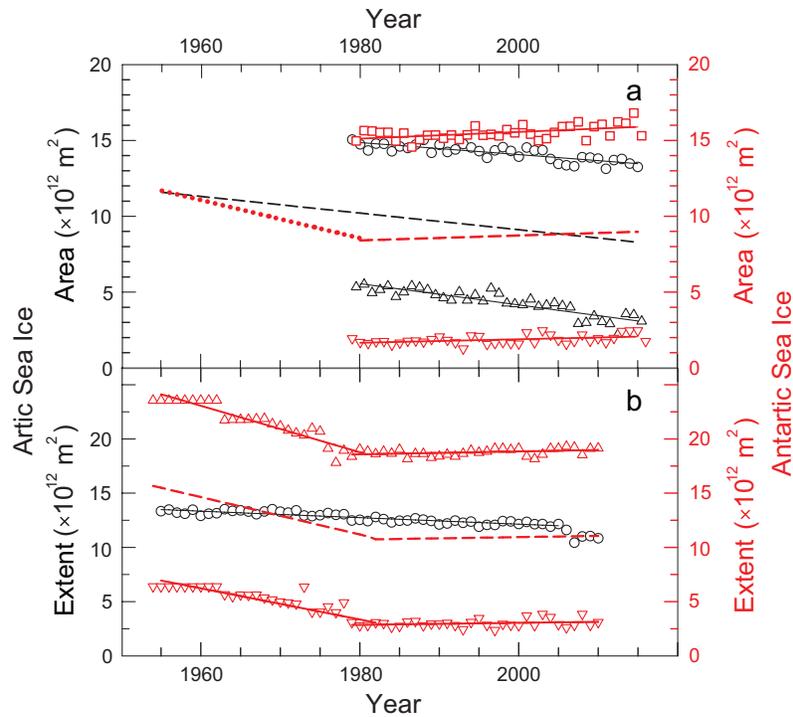

**Figure 1.** Melting of Arctic (black) and Antarctic sea ice (red) in the years 1955 to 2015. (a) Sea ice area and (b) sea ice extent; experimental points taken from literature and calculated lines; the dashed lines are the annual mean values of the summer and winter data; see text; after (16–21).

The effect of GHG gases for the global warming stems from the spectral properties of these gases. $CO_2$ and the two gases $CH_4$ and $N_2O$ display well-known absorption bands in the far infrared (FIR) for the thermal emission of the terrestrial surface leading in combination with atmospheric water to the well-known greenhouse effect (22). Water is most important because of its broad absorption features in the FIR. The situation is well documented by satellite measurements (23,24). Those data also show that the FIR bands of carbon dioxide and other gases in turn give rise to thermal radiation that is weaker because of the lower temperature in the atmosphere (compare Stefan-Boltzmann law) (25).

Fig. 2 presents an overview of the situation in the FIR with a spectral resolution of 10 cm$^{-1}$. The combined absorption of $H_2O$ and $CO_2$ in the medium part of the atmosphere (layer 2, see below) is plotted versus frequency in units of wave numbers in the range 0 – 3100 cm$^{-1}$ (black line, left hand ordinate scale). The concentrations refer to 1880. The total absorption, $A = 1$, in the centers of the narrow $CO_2$ bands at 667 and 2349 cm$^{-1}$ is readily seen (26). The strong absorption of $H_2O$ (1270 to 2150 and above 2800cm$^{-1}$) is noteworthy. The regions with smaller water absorption ($A < 1$) represent the spectral windows of the atmosphere for the thermal emission of the terrestrial surface into the universe (27). The combined absorption of $N_2O$ around 589, 1285 and 2220 cm$^{-1}$ (28) and of $CH_4$ around 1300 cm$^{-1}$ (26,28) is shown by the red curve in Fig. 2 (right hand ordinate scale). To show more details the red ordinate scale is down-shifted by 0.05. The concentrations again refer to the year 1880 (see below). The strong overlap with water absorption should be noted (see black curve). Comparison of the red and the black curves in the Fig. directly suggests that dinitrogen-oxide and methane play a minor role for the greenhouse effect of the atmosphere.

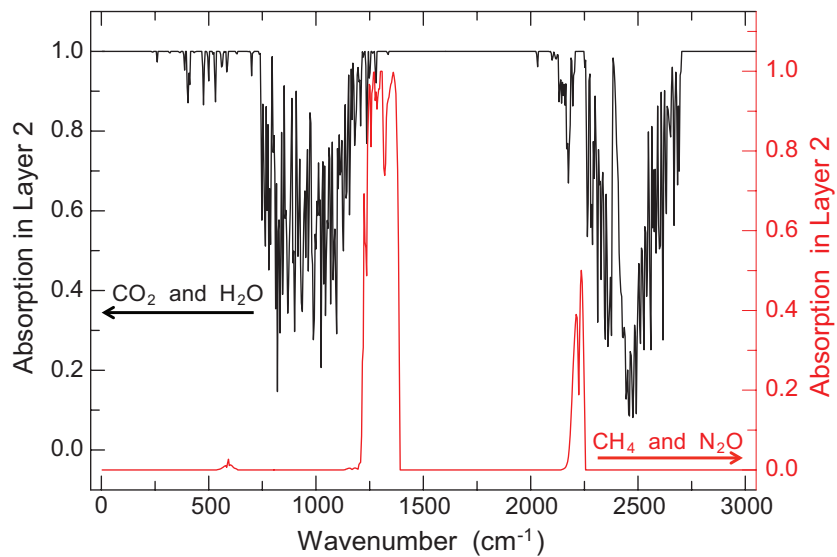

**Figure 2.** Spectral overview of the absorption of greenhouse gases in the medium part of the atmosphere (layer 2): water and carbon dioxide (black line, left hand ordinate scale); methane and dinitrogen oxide (red curve, right hand ordinate scale); the concentrations refer to 1880; see text.

We have carefully measured the FIR transmission of $CH_4$, $N_2O$ and of $CO_2$ in the wave number range 400 to 4000cm$^{-1}$. The water data are taken from an online database (29). Our experimental system was a Tensor 27 Fourier-transform infrared spectrometer (Bruker Optics) with a spectral resolution of 0.4cm$^{-1}$. Optically thick samples were investigated using an absorption cell of 10cm with KBr windows. To remove impurities, the sample was filled and evacuated several times with high purity gas. The total pressure with $N_2$ buffer gas was kept at 2 bar for measurements in the range 220 – 295 K. The temperature was controlled by a cryostat and directly measured in the sample. Part of the $CO_2$ data were published recently (30).

THEORETICAL MODEL:

The change of the mean global temperature $\Delta T_{surf}$ of the terrestrial surface is calculated by a 1-dimensional model as a function of the concentrations $x_i$ of the GHG (subscripts $M$=$CH_4$, $N$=$N_2O$, $C$=$CO_2$) and the albedo. The model is based on the spectroscopic properties of the gases and empirical results, e.g. the albedo factor in 1880 and the solar radiation input. The atmosphere is represented by four layers with mean temperatures $T_j$ (j = 1 – 4). The approach is an extension of the well-known 2-layer model for the greenhouse effect (30-32). We consider the 1-dimensional treatment quite satisfactory in the sense of a Taylor expansion of the (unknown) function $\Delta T_{surf}$ depending on a variety of local and seasonal parameters and GHG concentrations, where second and higher order terms are neglected. Retaining only the first order terms of the expansion, the problem is linearized and averaging over the manifold of parameters can be interchanged leading to a 1-D treatment. We have also considered an analogous 4-layer model arriving at similar results. It is concluded that our 5-layer approach provides realistic data on the mean global temperature change (see below).

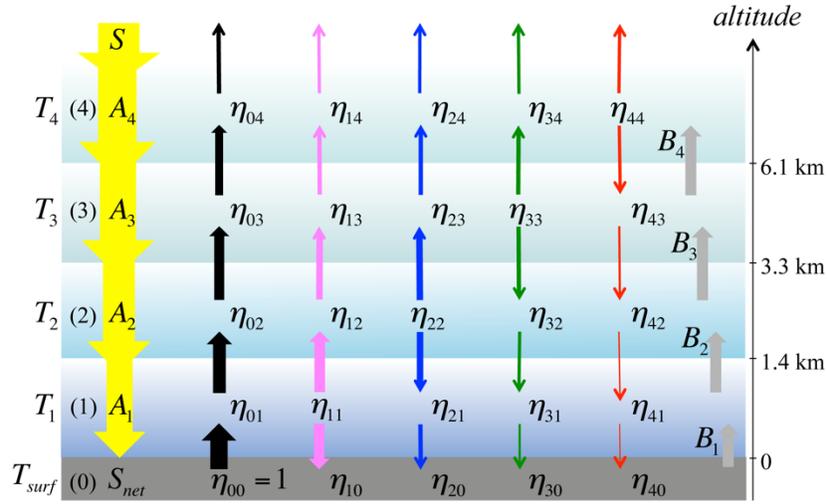

**Figure 3.** 5-level model for the global greenhouse effect with the terrestrial surface (0) and four atmospheric layers (1) – (4). The solar input is indicated by yellow arrows with parameters $S$ (total solar input), net surface input $S_{net}$ and layer inputs $A_1$ to $A_4$ (atmosphere). The thermal emission of the layers is illustrated by arrows. The $\eta_{ij}$ ($i,j = 1 - 4$) denote the spectral efficiencies of the respective layers $j$. The surface temperature is $T_{surf}$, while $T_1$ to $T_4$ are the mean temperatures of the other layers. The heat transport by non-radiative mechanisms is represented by gray arrows (parameters $B_1 - B_4$); see text.

Fig. 3 illustrates the model. Assuming equal absorption of the four atmospheric parts for $CO_2$, the thickness of the layers is estimated from a barometric formula to be 1.4, 1.9, and 2.8 km (layer 1 to 3, respectively), while the top layer extends above 6.1 km into the universe. The solar input is indicated in the Fig. by yellow arrows with parameter $S$ (total solar input) and $A_j$ for the atmospheric parts, while $A_{in} = \Sigma A_j$ denotes the total amount. The net solar input to the surface is $S_{net} = S \cdot (1 - albedo) - A_{in}$. The thermal emission of the layers is illustrated by arrows. The heat transport between the layers by non-radiative processes is represented by grey arrows (parameters $B_j$). In order to simplify the equations discussed just below a formal parameter $B_5 = 0$ is introduced. The spectral efficiencies of the respective layers, $\eta_{ij}$ ($i,j = 0 - 4$), are indicated that are calculated from the measured absorption properties of the greenhouse gases (including water) for the respective temperatures $T_j$ without a fitting parameter. The temperature $T_i$ of the emitting surface ($i = 0$) or atmospheric parts ($i = 1 – 4$) also comes in, since the spectral intensity distribution of the emission is temperature-dependent (compare Planck's formula) (25). The molecular number densities in the layers are calculated from a generalized barometric law. Quasi-equilibrium between input and output of the individual layers leads to the following equations for the surface (Eq. 1) and for the atmospheric layers $j = 1 – 4$ (Eq. 2):

$$\sum_{i=1}^{4} \sigma \cdot \eta_{i0} T_i^4 = \sigma \cdot T_{surf}^4 - S \cdot (1 - albedo) + A_{in} + B_1, \qquad (1)$$

$$\sigma \cdot \eta_{0j} T_{surf}^4 + \sum_{i=1}^{4} \sigma \cdot \eta_{ij} T_j^4 = 3\sigma \cdot \eta_{jj} T_j^4 - A_j - B_j + B_{j+1}, \qquad (2)$$

The surface is represented by a black radiator with intensity $\sigma T_{surf}^4$ (Stefan-Boltzmann constant $\sigma$) while the atmospheric parts are treated as selective thermal emitters. Transport phenomena between the layers are included by parameters $B_j$ but are found to have little quantitative effect on the surface warming (30). The temperatures $T_j$ are self-consistently determined from thermal quasi-equilibrium between the solar input and the thermal emissions of the surface and atmospheric layers; i.e. energy conservation (per unit time) is maintained for input and output radiation into the universe. The quantities $S = 343.3$ W/m² (total solar input), albedo $= 0.300$ and the surface temperature $T_{surf} = 288.0$ K are kept constant for the year 1880 as empirical facts. To evaluate the GHG effect, the albedo is constant in 1880 - 2015. To include the MSI contribution, the total albedo = albedo + $\Delta$albedo is allowed to vary,

The absorption $A_{in}$ of the atmosphere is chosen to maintain $T_{surf}$ for 1880, requiring values of $A_j$ in the range 20 to 40 W/m² (total $A_{in} = 105.3$ W/m²). The water content (integrated molecular number density) in the atmosphere is taken to be $9.2 \cdot 10^{26}$ m$^{-2}$ in 1880 for saturated vapor density and a linear temperature gradient in the atmosphere of 6.5 K/km. The amount of water follows the average layer temperatures depending on the GHG concentrations of the individual year. An enhancement of $\Delta T_{surf}$ results by a factor of approximately 1.3 compared to constant water densities in the layers. The number is somewhat smaller than reported by Wang *et al.* (28).

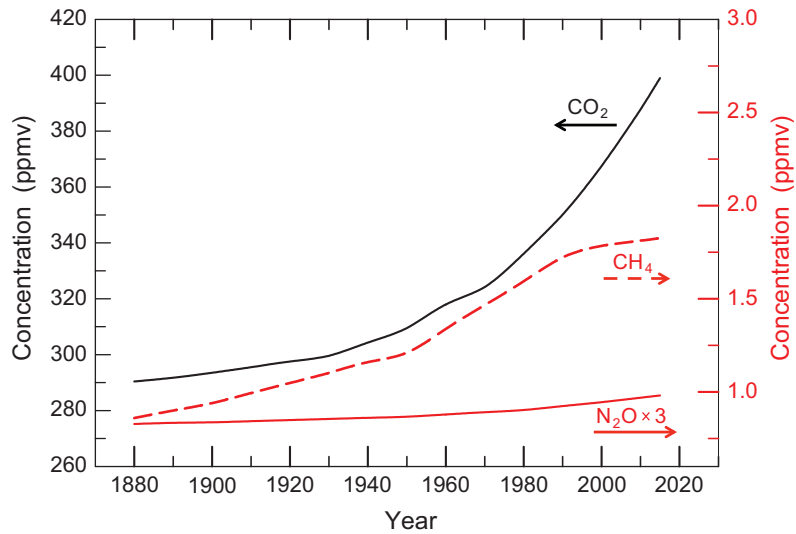

**Figure 4.** Reported abundance of greenhouse gases in the atmosphere in the years 1880 to 2015. Concentrations of $CO_2$ (black, left ordinate scale), $CH_4$ and $N_2O$ (red dashed and red solid curves, respectively; right ordinate scale). The $N_2O$ data are multiplied by a factor of 3 for better visibility; after (33).

The reported concentrations $x_C$, $x_M$ and $x_N$ of the GHG - used in our computations - are plotted in Fig. 4 for the years 1880 to 2015 (33,34). The concentrations in 1880 are $x_C$ = 290.4 ppmv, $x_N$ = 0.3 ppmv and $x_M$ = 1.6 ppmv. Note the monotonic increase of $CO_2$ from 290 to 399 ppmv (black curve, left hand ordinate scale). For methane the values are considerably smaller (dashed red line, right hand ordinate scale). For dinitrogen oxide $3 \cdot x_N$ is plotted in the Fig. (solid red curve, rhs ordinate scale). The contribution of ozone (not shown in Fig. 4) is included in our computations for the top layer leading to full absorption in the interval 1000 to 1072 cm$^{-1}$.

The numerical solution of equations shown above delivers the mean temperatures of respectively 285.2, 278.1, 268.3 and 240.1 K, for layers 1 to 4. We have shown recently that the specific parameter values of $A_j$ and $B_j$ vary the layer temperatures of the atmosphere to some extent but have little influence on the concentration dependence of the surface temperature (30).

RESULTS AND DISCUSSION:

We now discuss the results on global warming. The rise $\Delta T_{meas}$ of the reported mean surface temperature (35) is depicted in Fig. 5 (grey circles, lhs ordinate scale). Up-shifting the GISS-data (35) by 0.15 K the value for 1880 is arbitrarily set to zero. An increase of $1.0 \pm 0.1$ K is indicated for the years 1880 to 2015. It is interesting to see the non-monotonic behavior with a pronounced decrease from 1880 to 1910 and again from 1940 to 1950. Quite obviously, the global temperature does not follow the rising GHG abundance in the same intervals. Other factors, not investigated in the present paper, notably influence the surface temperature.

The causal relationship between the GHG growth and the global surface temperature as evaluated from our model is also presented in Fig. 5 (plotted curves, rhs ordinate scale). The rise of the surface temperature was calculated for every year from the respective GHG concentrations. The result for constant *albedo* = 0.300 is shown by the green curve. A temperature rise of $\Delta T_{surf} = 0.256$ K is computed for the period 1880 - 2015 that originates mostly from the concentration increase of $CO_2$ from 290 to 399 ppmV (including water enhancement). The contributions of $CH_4$ of 46 mK, considerably smaller by a factor of 5, and that of $N_2O$ of only 1.6 mK are included. The simultaneous global warming of the atmosphere until 2015 is calculated to be 0.20, 0.15, 0.09 and

0.05 K for layers 1 to 4, respectively. We conclude that the FIR properties of the growing GHG in the atmosphere only generate a moderate contribution for the rising surface temperature.

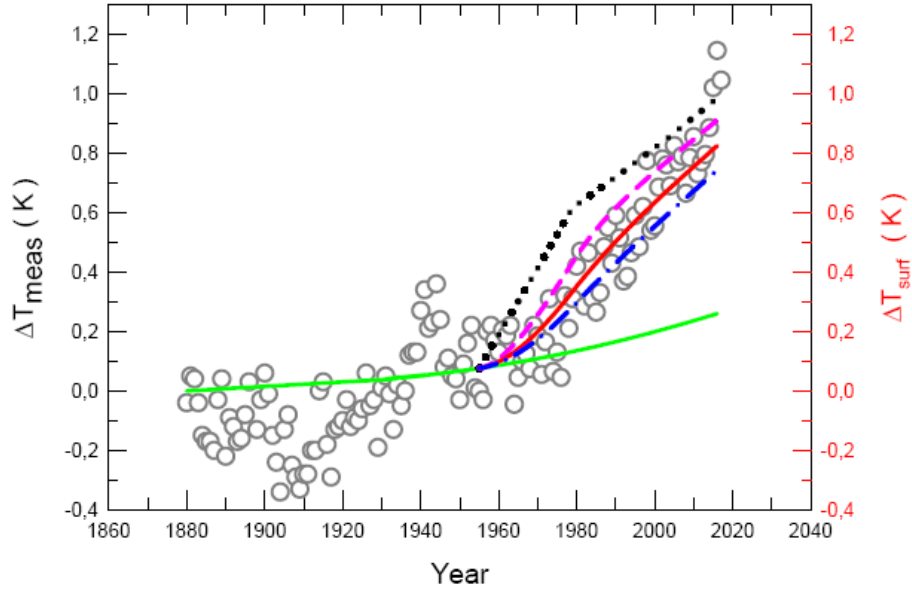

**Figure 5.** Reported change of the mean surface temperature $\Delta T_{meas}$ (35) (open circles, left hand ordinate) and calculated results $\Delta T_{surf}$ for the past 135 years (curves, right hand ordinate). The values of reference (35) are up-shifted by 0.15 K so that the mean temperature change in 1880 is approximately equal to zero. The rise by ~ 1.0 K until 2015 is noteworthy. For constant *albedo* = 0.300 the computed temperature change $\Delta T_{surf}$ derived from the spectroscopic properties of the GHG including water is shown by the green curve. For 1955 – 2015 the combined contribution of the GHG and of sea ice melting is shown for (i) instantaneous response of the global system (dotted black line); (ii) delayed equilibration of the global temperature with relaxation time $\tau$ = 10 yr (dashed pink curve), (iii) $\tau$ = 20 yr (solid red line) and (iv) $\tau$ = 30 yr (dash-dotted blue curve).

The important role of the polar sea ice for the global warming as revealed by our investigation is discussed now. The data presented in Fig. 1 indicate considerable MSI in the years 1955 to 2015. The data are used as empirical facts in our computations allowing an estimate of the corresponding decrease of the back-scattering of solar light from the pole regions into the universe. Details are discussed in the Appendix. The resulting lowering of the albedo factor is estimated to be $-2.9 \times 10^{-3}$ if instantaneous response of the global system to the changes in the polar regions is assumed. The corresponding warming of the surface as suggested from a solution of Eqs. 1 and 2 would amount to 0.99 K (dotted black line in Fig. 5). A steep temperature increase results for 1955 to 2015 that differs from the reported rise of the global surface temperature (compare experimental points in Fig. 5).

For a possible explanation of the deviation we recall that a prompt response of the mean global temperature to changes in the Arctic and Antarctic should not be expected. In fact, the well-known differences between equator and pole areas indicate a rather slow heat transport. As a consequence, we extend our theoretical model and tentatively incorporate a delayed response of the global system to changes of the net solar input in the pole regions. A simple relaxation ansatz is made with time constant $\tau$. Details are discussed in the Appendix. The results for relaxation times of $\tau =$ 10 yr (dashed pink line), 20 yr (solid red curve) and 30 yr (dash-dotted blue line) are presented in Fig. 5. The delayed response lowers the MSI increase until 2015 arriving at $\Delta T_{surf} = 0.91$ K for $\tau = 10$ yr, 0.82 K ($\tau = 20$ yr) and 0.74 K ($\tau = 30$ yr), respectively. It is felt that a time constant of $\approx$ 20 years (red curve in Fig. 5) accounts fairly well for the equilibration by heat transfer between the pole areas and the global system. Nevertheless, some room seems to be left for other contributions.

We emphasize that the simple 1-dimensional model of Eqs. 1 and 2 appears reliable. Tentatively we have extended our computations to a 2-dimensional approach incorporating the distinct variations of the solar input according to the geographical latitudes of the earth, arriving at similar results (deviations of several % compared to the 1-dimensional model; data not shown). Because of the various simplifications the accuracy of the 5-layer model of Eqs. 1 and 2 is believed to be ± 10%. The estimate of the albedo decrease by MSI may be accurate to ± 15 % for 1955 – 1979, only, while the data since 1979 are believed to be more accurate. Other factors affecting the albedo value and/or different mechanisms discussed in the literature (15,22,36) are expected to influence the global warming. It also was recognized recently that black carbon in the atmosphere is relevant, much more important than the concentration changes of ozone, methane and nitrous oxide (12).

Several authors discussed the feedback of surface reflectivity on climate (13,37–40). In this context it is interesting to compare our results with data of Flanner *et al.* (13), who quantified the albedo feedback of snow in the northern hemisphere (NH) between 1979 and 2008. Their result for the NH cryosphere albedo feedback was $\Delta F_{cryo}/\Delta T_{surf} = 0.62$ (0.3 – 1.1) W/(m²K), notably larger than estimated from various climate models (13). Our values for the albedo change and surface temperature rise for this period are respectively -5.84 x 10$^{-4}$ and 0.144 K leading to a similar value of $\Delta F_{ice}/\Delta T_{surf} = 0.68$ W/(m²K). The question remains about the origin of the melting of polar sea ice (41). Several authors point out that it is partly due to natural reasons (36,42). There is evidence for climate changes in past centuries evolving on time scales of hundreds of years that were accompanied by distinct changes of the Arctic sea ice (43–45). As to anthropogenic factors, the present authors wish to point to air pollution as a probable reason for the disappearance of sea ice and glaciers (41,46), possibly more relevant than the temperature rise, directly caused by the IR properties of the GHG. Some authors considered a GHG-contribution to the ice melting in the Arctic (47). Thus, part of the MSI-temperature effect of ~0.55 K may originate from an enhancement mechanism via the GHG.

CONCLUSIONS:

In conclusion we wish to say that we have performed a study of the infrared properties of carbon dioxide, methane, dinitrogen-oxide and water to estimate their contribution to the global warming in 1880 – 2015. Our results suggest that the IR properties of the $CO_2$ are responsible for ~ 20% of the mean temperature increase of the surface and notably less for $CH_4$ and $N_2O$. On the other hand, our analysis of existing data for the melting of polar sea ice reveals its dominant role for the global warming. Delayed response of the global system to the changing surface reflectivity with relaxation time ~ 20 years and a surface temperature rise since 1955 by ~0.8 K are calculated by our model.

APPENDIX: ESTIMATE OF THE ALBEDO DECREASE BY MSI

The data of Fig. 1 indicate a loss of Arctic sea ice area of 3.29 x $10^{12}$ m$^2$ for the years 1955 to 2015. For the Antarctic the behavior is non-monotonic with a decrease of 3.38 x $10^{12}$ m$^2$ in the same period. Melting replaces snow-covered sea ice by a water surface reducing the backscattering by a factor of ≈ 0.5 (14,48). Averaging over the daily and seasonal changes of the solar input for the respective areas around a northern (southern) latitude of ≈ 82° (≈ 63°) (21,49) and also including a transmission of 0.49 of the atmosphere for the sun input to the pole regions leads to a reduction factor of 0.112 for the solar intensity of 1367 W/m$^2$. The decrease of the terrestrial back-reflection into the universe thus arrives at $\Delta P_{refl}$ = -5.01 x $10^{14}$ W in 2015. Relative to the total solar input of the earth, $P_{tot}$=1.75 x $10^{17}$ W, we estimate a relative change of $G = \Delta P_{refl}/ P_{tot}$ = -2.9 x $10^{-3}$. For instantaneous response of the global system this number would correspond to a relative albedo decrease of ≈ -0.97 % in 2015. $G$ can be treated as a known function of time according to the MSI data of Fig. 1. For delayed response of the global system leading to an albedo decrease $\Delta albedo(t)$, we introduce the simple relaxation ansatz of Eq. 3 with a relaxation time $\tau$ of the global system:

$$\frac{d}{dt}\Delta albedo + \frac{\Delta albedo}{\tau} = \frac{G(t)}{\tau}.\tag{3}$$

The global albedo = 0.300 + $\Delta albedo(t)$ is now a time-dependent parameter because of the changing backscattering of the earth via MSI in 1955 to 2015. Eq. 3 is readily solved for $\Delta albedo(t = 1955) = 0$ and decreases to $-2.25 \times 10^{-3}$ in 2015 for $\tau = 20$ yr ($-2.6 \times 10^{-3}$ for $\tau = 10$ yr; $-1.9 \times 10^{-3}$ for $\tau = 30$ yr). The use of Eq. 3 in context with our theoretical model is supported by the approximately constant ratio $\Delta albedo / \Delta T_{surf}$ delivered by Eqs. 1 and 2; i.e. approximately linear response of the global system to small albedo changes. We mention that the albedo decline by MSI discussed here is lacking direct experimental verification. Changes of the global albedo of opposite signs (± 0.02) were reported in the years 1985 – 1997 and 1997 – 2004 (50,51) (but not supported by comparison with the reported temperature data), and on the contrary that the global albedo remained fairly constant during the past decades (52). We propose that the pronounced rise $\Delta T_{meas}$ of the surface temperature in Fig. 5 in 1970 – 2015 represents some experimental support for our estimate of the albedo decrease.